\documentclass{JHEP3}

\usepackage{epsfig}
\usepackage{amsmath}

\newcommand{\be}{\begin{equation}}
\newcommand{\ee}{\end{equation}}
\newcommand{\bea}{\begin{eqnarray}}
\newcommand{\eea}{\end{eqnarray}}

\title{The spatial string tension in the
deconfined phase of three dimensional QCD in the large N limit}
\author{J. Kiskis\\
Department of Physics, University of California,
Davis, CA 95616.\\E-mail: \email{jekiskis@ucdavis.edu}}
\author{R. Narayanan\\
Department of Physics, Florida International University, Miami,
FL 33199.
\\E-mail: \email{rajamani.narayanan@fiu.edu}}

\abstract{
We numerically compute the spatial string tension in the
deconfined phase of three dimensional QCD in the large N limit.
Our results clearly show that the 
string tension grows linearly with
temperature.
}

\keywords{
Large $N$ QCD, Deconfined phase, Spatial string tension}

\preprint{}

\begin{document}

Three dimensional QCD on a symmetric torus is considered
in the large N limit. This theory is known to be in the
deconfined phase for $T_1 < T < T_2$~\cite{Narayanan:2007ug}. 
$T_1$ is the
temperature at which one of the three $Z_N$ symmetries
is broken and $T_2$ is the temperature at which two of
the three $Z_N$ symmetries are broken\footnote{$T_2$ can be
taken to infinity by working on an asymmetric torus with one
short (temperature) direction and two long (spatial) directions.}.

At leading order in $1/N$, 
the string tension is independent of the temperature in the
confined phase ($T < T_1$), and it
was computed using reduction~\cite{Kiskis:2008ah}.
Since the $Z_N$ symmetry is broken only in one direction
in the deconfined phase, spatial Wilson
loops in the plane perpendicular to the broken direction
are expected to show confining behavior.
This has been numerically observed 
in the deconfined phase of SU(2) Yang-Mills
theory in four dimensions~\cite{Bali:1993tz}.

We use the standard Wilson gauge action,
\be
S=bN\sum_{{\bf n},\hat{\bf i}\ne \hat{\bf j}} {\rm Re\ Tr}
U_i ({\bf n}) U_j ({\bf n}+\hat {\bf i}) 
U_i^\dagger ({\bf n}+\hat {\bf j}) 
U_j^\dagger ({\bf n}),
\ee
on a $L^3$ lattice
with periodic boundary conditions. 
The 't Hooft limit is obtained by
keeping the
coupling $b=\frac{1}{g^2N}$ fixed as we take $N\to\infty$. We use
the tadpole improved lattice coupling~\cite{Lepage:1996jw},
\be
b_I = b e(b) = \frac{1}{6NL^3}\langle S\rangle,
\ee
instead of the lattice coupling $b$.
We approach the continuum
limit in the deconfined phase
by taking $b_I\to\infty$ and $L\to\infty$
keeping the physical temperature, $T=\frac{b_I}{L}$ fixed. 
The upper and lower limits of the deconfined phase
have been numerical estimated~\cite{Narayanan:2007ug}
to be 
\bea
T_1 &= \lim_{L\to\infty} \frac{{b_I}_1(L)}{L} &= 0.160(13)\cr
T_2 &= \lim_{L\to\infty} \frac{{b_I}_2(L)}{L} &= 0.263(19).
\label{trantemp}
\eea

{\tiny
\TABLE[htp]{
\begin{tabular}
{ccclllllll}
$L$ & $N$ & $b$ & \ \ \ \ \ $e$ & \ \ \ \ \ $\langle P_1\rangle$ & \ \ \ \ \ $\langle P_2\rangle$
& \ \ \ \ \ $\langle P_3\rangle$ & $\ \ \ \ \ C_0$ & $\ \ \ \ \ C_1$ & $\ \ \ \ \ \sqrt\Sigma$\cr
\hline
  $4$ & $47$ & $ 0.900$ & $0.801529(17)$ & $ 0.45386(22)$ & $ 0.49813( 4)$ & $ 0.49927( 2)$ & $-0.0532(15)$ & $ 0.1586(16)$ & $ 0.28283(49)$\\
  $4$ & $47$ & $ 1.000$ & $0.823036(15)$ & $ 0.42530(21)$ & $ 0.49771( 6)$ & $ 0.49922( 2)$ & $-0.0420(13)$ & $ 0.1344(13)$ & $ 0.26521(40)$\\
  $4$ & $47$ & $ 1.100$ & $0.840264(14)$ & $ 0.40359(25)$ & $ 0.49479(14)$ & $ 0.49823( 5)$ & $-0.0306(10)$ & $ 0.1131(11)$ & $ 0.25051(36)$\\
  $4$ & $59$ & $ 0.900$ & $0.801499(14)$ & $ 0.45513(32)$ & $ 0.49750(11)$ & $ 0.49896( 5)$ & $-0.05167(15)$ & $ 0.1569(15)$ & $ 0.28327(42)$\\
  $4$ & $59$ & $ 1.000$ & $0.823003(13)$ & $ 0.42379(15)$ & $ 0.49861( 3)$ & $ 0.49947( 1)$ & $-0.0595(16)$ & $ 0.1520(18)$ & $ 0.26114(63)$\\
  $4$ & $59$ & $ 1.100$ & $0.840220(11)$ & $ 0.40016(17)$ & $ 0.49744( 6)$ & $ 0.49922( 2)$ & $-0.0455(12)$ & $ 0.1274(13)$ & $ 0.24827(52)$\\
  $5$ & $59$ & $ 1.000$ & $0.822623( 9)$ & $ 0.49381(12)$ & $ 0.49776( 4)$ & $ 0.49907( 2)$ & $-0.05458(53)$ & $ 0.15586(59)$ & $ 0.24446(21)$\\
  $5$ & $59$ & $ 1.125$ & $0.843795( 8)$ & $ 0.45277(22)$ & $ 0.49831( 6)$ & $ 0.49928( 3)$ & $-0.05148(56)$ & $ 0.13954(57)$ & $ 0.22553(22)$\\
  $5$ & $59$ & $ 1.250$ & $0.860343( 7)$ & $ 0.42594(15)$ & $ 0.49865( 3)$ & $ 0.49955( 1)$ & $-0.04660(43)$ & $ 0.12423(46)$ & $ 0.21094(21)$\\
  $5$ & $59$ & $ 1.375$ & $0.873694( 6)$ & $ 0.40449(16)$ & $ 0.49764( 5)$ & $ 0.49915( 2)$ & $-0.03857(37)$ & $ 0.10847(40)$ & $ 0.19909(19)$\\
  $6$ & $59$ & $ 1.200$ & $0.854075( 5)$ & $ 0.48168(19)$ & $ 0.49848( 4)$ & $ 0.49947( 1)$ & $-0.05681(30)$ & $ 0.14171(32)$ & $ 0.20503(13)$\\
  $6$ & $59$ & $ 1.350$ & $0.871170( 5)$ & $ 0.45220(14)$ & $ 0.49875( 2)$ & $ 0.49952( 1)$ & $-0.04668(23)$ & $ 0.12105(25)$ & $ 0.19082(12)$\\
  $6$ & $59$ & $ 1.500$ & $0.884638( 4)$ & $ 0.42848(15)$ & $ 0.49852( 3)$ & $ 0.49935( 2)$ & $-0.04140(19)$ & $ 0.10774(21)$ & $ 0.17810(12)$\\
  $6$ & $59$ & $ 1.650$ & $0.895545( 4)$ & $ 0.40783(14)$ & $ 0.49800( 3)$ & $ 0.49922( 2)$ & $-0.03671(17)$ & $ 0.09652(19)$ & $ 0.16788(11)$\\
  $7$ & $59$ & $ 1.325$ & $0.868531( 4)$ & $ 0.49794( 4)$ & $ 0.49889( 2)$ & $ 0.49950( 1)$ & $-0.04721(15)$ & $ 0.12403(17)$ & $ 0.19006( 8)$\\
  $7$ & $59$ & $ 1.400$ & $0.875947( 3)$ & $ 0.48714(17)$ & $ 0.49713( 5)$ & $ 0.49907( 2)$ & $-0.04274(15)$ & $ 0.11540(16)$ & $ 0.18275( 8)$\\
  $7$ & $59$ & $ 1.575$ & $0.890335( 3)$ & $ 0.45255(11)$ & $ 0.49864( 3)$ & $ 0.49951( 1)$ & $-0.03855(12)$ & $ 0.10265(14)$ & $ 0.16857( 7)$\\
  $7$ & $59$ & $ 1.750$ & $0.901710( 3)$ & $ 0.43089(13)$ & $ 0.49850( 3)$ & $ 0.49947( 1)$ & $-0.03740(10)$ & $ 0.09472(11)$ & $ 0.15605( 7)$\\
\hline
\end{tabular}
\caption{\label{tab1}
The various parameters ($L$, $N$ and $b$), the average value of the plaquette, $e$,
Polyakov loop order parameter in the three directions and the three parameters in the fit
of the static potential. Each set has $800$ measurements. All errors
are obtained using single elimination jackknife.}}}

We used $L=4,5,6,7$ and $N=59$ to extract the behavior of
the string tension as a function of the temperature in the
deconfined phase. We also used $L=4$ and $N=47$ to verify
that $N=59$ was large enough for us to ignore finite $N$ effects.
The complete list of couplings used in the numerical analysis
can be found in Table~\ref{tab1}, and the full
range from $T_1$ and $T_2$ has been covered. A plot of
the average energy $e(b)$ listed in Table~\ref{tab1} is plotted
in Figure~\ref{energy} along with a fit that includes a $\frac{1}{b}$
and a $\frac{1}{b^2}$ term. The tadpole improved coupling
removes some large lattice spacing effects that are 
common
to all observables.

\FIGURE[htp]{
\epsfig{file=energy.eps, height=4in }
\caption{Plot of $e(b)$ as a function of $b$ for all the data
sets in Table~\ref{tab1}.}
\label{energy}}

We used an
order parameter based on the Polyakov loop and defined 
by~\cite{Bhanot:1982sh}
\bea
\bar P_i &=& \left < P_i \right > \cr
P_i &=& \frac{1}{2 L^2} \sum_{\bf n} ( 1 - \left | \frac{1}{N} 
Tr {\cal P}_i({\bf n})
\right |^2 ) \cr
{\cal P}_i({\bf n}) &=& \prod_{m=1}^{L} U_i({\bf n}+m\hat {\bf i}),
\eea
to ensure that we are in the deconfined phase.
The quantity $P_i$ takes values in the range $[0,0.5]$ on
any gauge field background, and we choose the 
directions $i=1,2,3$ such that
such that $P_1 < P_2 < P_3$. The deconfined phase corresponds
to $\bar P_1 < 0.5$ and $\bar P_2 = \bar P_3 = 0.5$ within errors.
The data presented in Table~\ref{tab1} and plotted
in Figure~\ref{poly} shows this to be the
case.

\FIGURE[htp]{
\epsfig{file=poly.eps, height=4in }
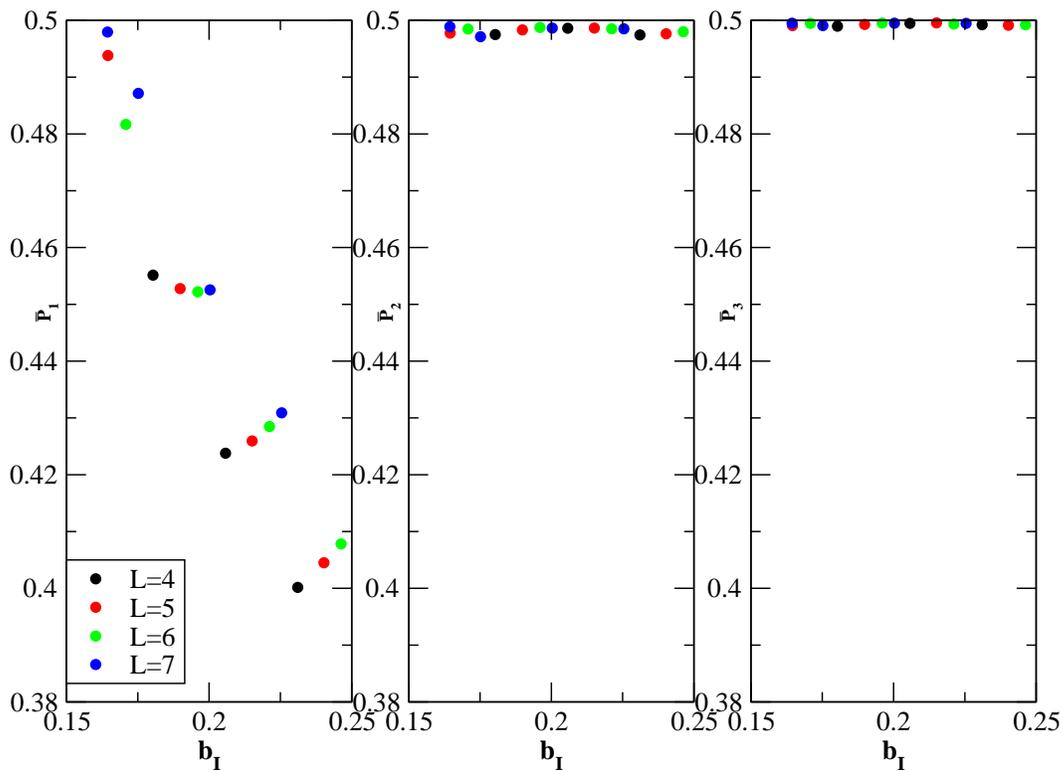
\caption{Plot of $\bar P_i$ as a function of $b$ for all the data
sets in Table~\ref{tab1} with $N=59$
showing that we are in the deconfined phase.}
\label{poly}}

We use computational techniques identical to the ones
described in~\cite{Kiskis:2008ah} to compute the static
potential in the deconfined phase of large N QCD for spatial Wilson loops.
As discussed in~\cite{Kiskis:2008ah},
our links in the $1-2$ plane are smeared while the links
in direction $3$ are not smeared. 
We have set the smearing factor to $f=0.1$ and the number
of smearing steps to $n=25$. We compute the expectation values of all
$L_2\times L_3$ loops, $W(L_2,L_3)$,
 in the $2-3$ (spatial) plane with $1\le L_2,L_3 \le 10$.
Keeping $L_2$ fixed, we fit
\be
\ln W(L_2,L_3) = -A -V(L_2)L_3.
\ee
Since $W(L_2,L_3)$ becomes small as $L_2L_3$ becomes big, we do not
use all values of $L_3$ for the fit. The range of $L_3$ is determined
such that the relative error of the fit per degree of freedom (number
of values of $L_3$ used in the fit) is less than $0.1\%$, and this
typically included all loops with $1 \le L_2L_3 \le 40$.
The static potential $V(L_2)$ is subsequently fit to
\be
V(L_2) = C_0 + \frac{C_1}{L_2} + \Sigma L_2.
\ee
The results for $C_0$, $C_1$ and $\sqrt{\Sigma}$ are shown in 
Table~\ref{tab1}. 
The performance of the fits for $L=4,5,6,7$ at $N=59$ can
be seen in the plots of the static potential 
in Figure~\ref{potential}.
A comparison of the values 
of $\sqrt{\Sigma}$ for $N=47$ and $N=59$ at $L=4$
for the three different values of $b$ shows that finite $N$ effects
are small at $N=59$.

\FIGURE[htp]{
\epsfig{file=potential.eps, height=4in }
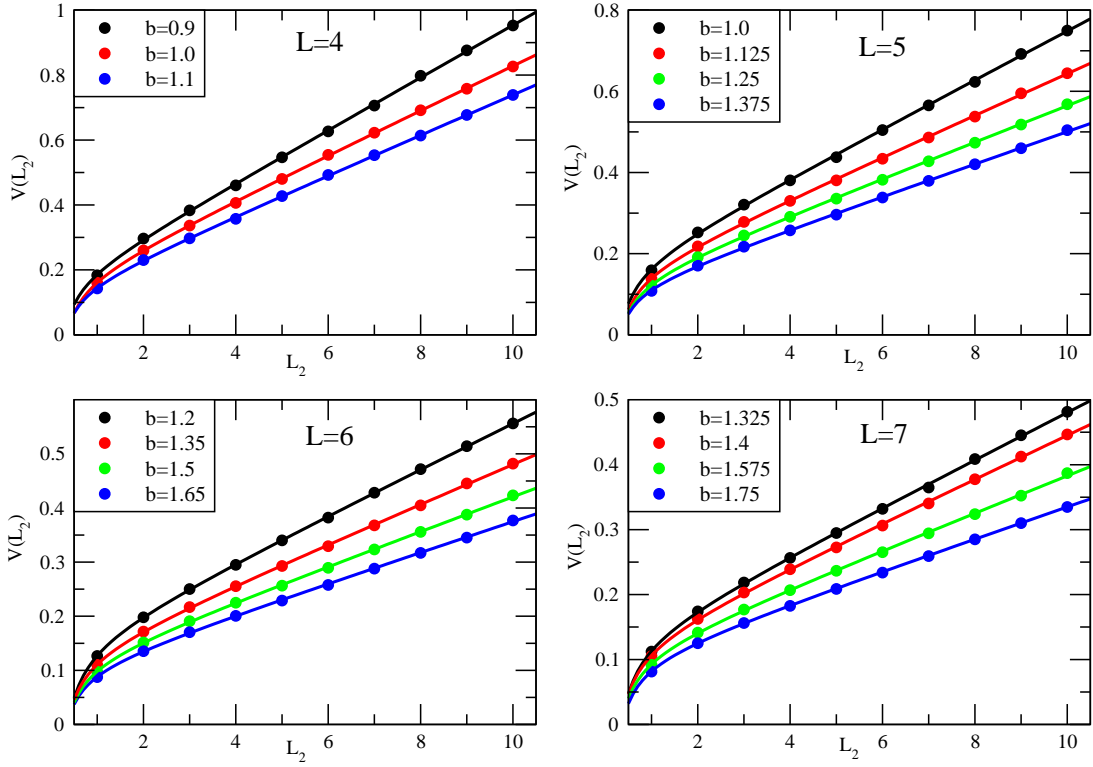
\caption{Plot of the static potential
as a function of the separation for all the data
sets in Table~\ref{tab1} with $N=59$.}
\label{potential}}

The last column in Table~\ref{tab1} provides
all the results obtained in this paper for the
 spatial string tension
at several values of the temperature in the
deconfined phase. If we use high temperature dimensional
reduction similar to the arguments in~\cite{Bali:1993tz},
the effective two dimensional coupling is
\be
b_2=bL.
\ee
If the spatial string tension in the deconfined phase
is equal to the string tension in the two dimensional
theory, then we expect~\cite{Gross:1980he}
\be
\Sigma = \frac{1}{4b_2} = \frac{1}{4bL}.\label{string2d}
\ee
A plot of $\Sigma$ versus $\frac{1}{bL}$ is shown in
Figure~\ref{stringkb}. The combined data at the four
different values of $L$ are fit to a straight line.
Considering that the value of $T$ is modest and that its value is not 
large,
the fit is surprisingly consistent with the prediction in 
(\ref{string2d}).
\FIGURE[htp]{
\epsfig{file=stringkb.eps, height=4in }
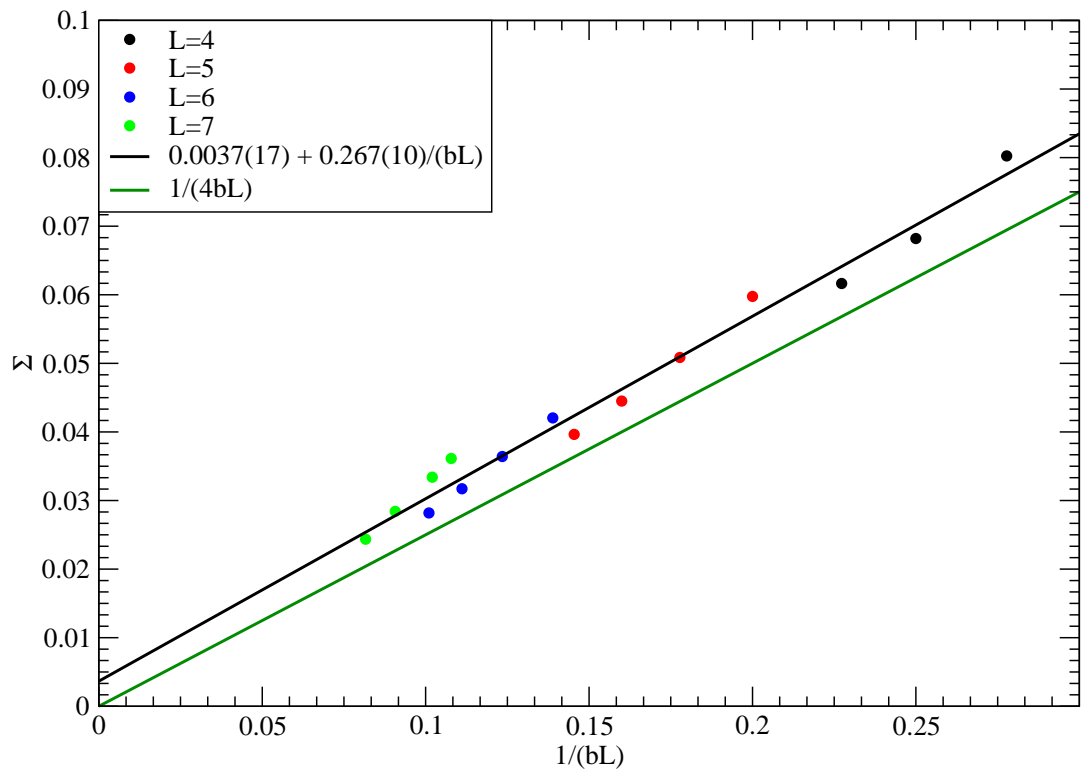
\caption{The spatial string tension 
as a function of the effective two
dimensional coupling, $\frac{1}{bL}$ in the deconfined phase.}
\label{stringkb}}

Another way to plot
Figure~\ref{stringkb} 
is to convert the axes to physical units (physical string
tension vs. physical temperature) and use tadpole improved
coupling. Such a plot is shown in Figure~\ref{string1}.
At the same physical $T$, larger $L$ corresponds to finer lattice 
spacing. 
Thus finite spacing effects appear as the scatter in $L$ in this plot,
and the fit coefficients are slightly
different from those in Figure~\ref{stringkb}.
\FIGURE[htp]{
\epsfig{file=string1.eps, height=4in }
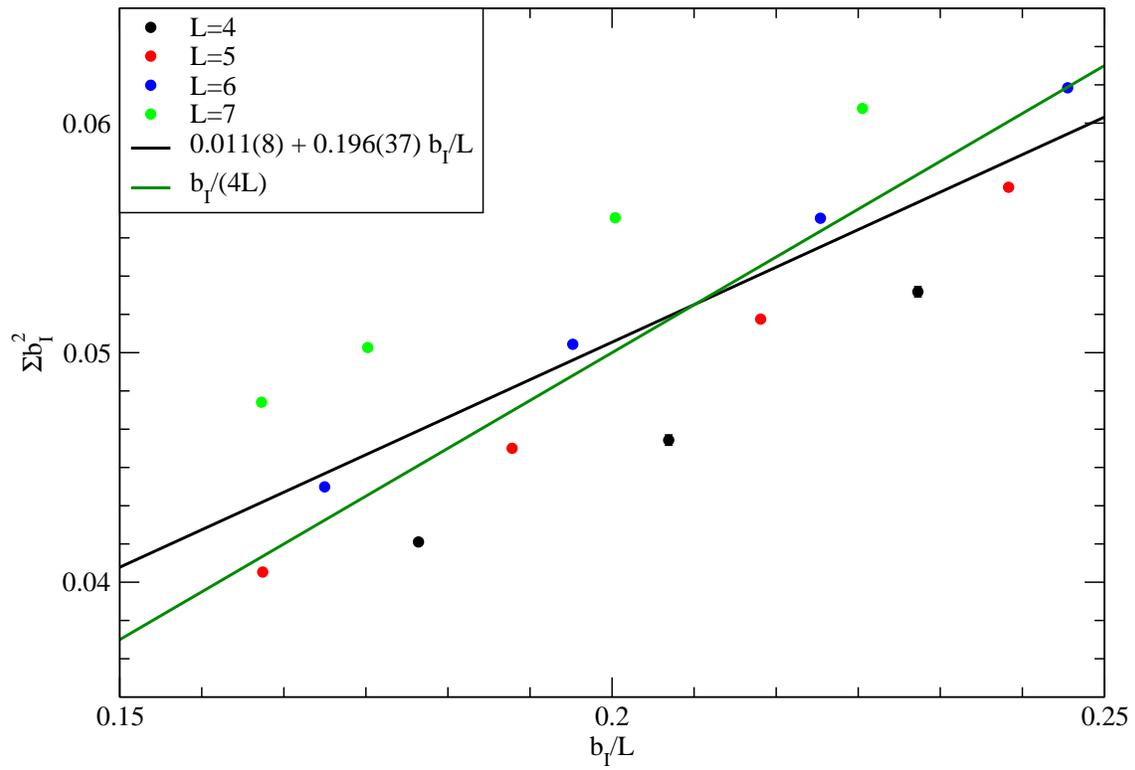
\caption{The dependence of the spatial string tension 
on the temperature in the deconfined phase.}
\label{string1}}

If we assume (\ref{string2d}) to be valid in the entire
region of the deconfined phase and use 
$\frac{1}{8\pi b^2}$~\cite{Karabali:1998yq} for
the string tension in the confined phase, continuity
of the string tension across the phase transition suggests
that $T_1=\frac{1}{2\pi}$ consistent with (\ref{trantemp}).

Alternatively, we can compare the $L=5$ results here and 
in~\cite{Kiskis:2008ah}. When 
the measured values for $\sqrt{\Sigma}$ at $T=0$ are extrapolated to 
$b=1.00$ we obtain 0.232(11), which compares favorably with the 
corresponding value in Table~\ref{tab1}.

If the spatial string tension is indeed continuous at $T_1$, then we 
have a picture for the $N=\infty$ limit in which the spatial and 
temporal string tensions are equal and independent of $T$ for $T<T_1$. 
At $T_1$, the temporal string tension falls to zero discontinuously, and 
the spatial string tension begins to grow linearly with $T$.

\acknowledgments

R.N. acknowledges partial support by the NSF under grant number
PHY-055375.

\end{document}